\documentclass[11pt,a4paper]{article}
\usepackage{vmargin}
\usepackage[T1]{fontenc}
\usepackage{epsfig}
\usepackage{psfig}
\usepackage{graphicx}
\usepackage{graphics}
\usepackage{xspace}
\usepackage{amssymb}
\usepackage{amsmath}
\usepackage{latexsym}
\usepackage{mathrsfs}
\newcommand{\diff}{{\rm\,d}}                    %simbolo di derivata totale
\newcommand{\ove}{\overline}                    %barra sopra lettera 
\def\fps@figure{htb}
\def\fps@table{htb}
\def\r{\mbox{\boldmath $r$}}

\def\p{\mbox{\boldmath $p$}}
\def\P{\mbox{\boldmath $P$}}
\def\q{\mbox{\boldmath $q$}}

\def\J{\mbox{\boldmath $J$}}

\def\ss{\mbox{\boldmath $\sigma$}}

\def\dd{\mbox{\boldmath $\nabla$}}

\def\r{\mbox{\boldmath $r$}}

\mathchardef\varepsilon="010F
\mathchardef\epsilon="0122
\mathchardef\theta="0123
\mathchardef\vartheta="0112
\begin{document}
\title{Relativistic corrections in $(\gamma,N)$ knockout reactions}
\author{A.~Meucci, C.~Giusti, and F.~D.~Pacati \\
\emph {Dipartimento di Fisica Nucleare e Teorica, Universit\`a di 
Pavia}\\
\emph {and Istituto Nazionale di Fisica Nucleare, Sezione di Pavia, Italy}}
%\date{\today}

\maketitle

%%%%%%%%%%%%%%%%%%%%%%%%%%%%%%%%%%%%%%%%%%
\begin{abstract}
%\vskip .5cm
We develop a fully relativistic DWIA model for photonuclear reactions 
using the relativistic mean field theory for the bound state and the 
Pauli reduction of the scattering state which is calculated from a relativistic
optical potential. Results for the $^{12}$C$(\gamma,p)$ and
$^{16}$O$(\gamma,p)$
differential cross sections and photon asymmetries are displayed in a
photon energy range between 60 and 257 MeV, and compared with 
nonrelativistic DWIA calculations.  
The effects of the spinor distortion and of the effective momentum approximation 
for the scattering state are discussed. The sensitivity of the model to
different prescriptions for the one-body current operator is investigated. The 
off-shell ambiguities are large in $(\gamma,p)$ calculations, and even larger
in $(\gamma,n)$ knockout.
\end{abstract}


%%%%%%%%%%%%%%%%%%%%%%%%%%%%%

\section{Introduction}

The analysis of ($\gamma,N$) reactions at photon energies above the giant
resonance was the object of a long debate concerning the mechanism of the
reaction (see e.g. Ref.~\cite{Oxford}). On the one hand, the fact that the 
experimental cross sections for proton emission can be easily fitted with a 
single particle wave function addresses to a direct knockout (DKO) 
mechanism~\cite{Boffi81}. 
On the other hand, the transitions with neutron emission, being of the same 
order of magnitude as those with proton emission, were considered as a clear 
indication of a quasi-deuteron reaction mechanism~\cite{Levinger,Schoch,Eden}.
A number of corrections were applied to the DKO model~\cite{Boffi84,Boffi85} in 
order to explain both ($\gamma,p$) and ($\gamma,n$) cross sections, but were 
unable to give a reasonable explanation of the data.

In recent years, the development of tagged photon facilities allowed to perform
experiments with high energy resolution and a clear separation of the different 
individual states of the residual nucleus.
A large number of experimental data was produced
at the electron microtron accelerator MAMI-A in Mainz and at the MAX-Laboratory 
in Lund (see e.g. Refs.~\cite{spring,miller,Abeele,Andersson,deBever,Branford}).

For the ($\gamma,p$) reaction the DKO mechanism represents a large part of the
measured cross sections for the low-lying states and in the photon energy range 
above the giant resonance and below the pion production threshold. The results, 
however, are very sensitive to the theoretical ingredients adopted for bound 
and scattering states~\cite{Boffi81,Benenti}. Moreover, various calculations 
in different theoretical approaches indicate that a prominent role is played 
by more complicated processes, like meson exchange currents (MEC) and 
multi-step processes due to nuclear correlations~\cite{Oxford,miller,Benenti}.
Nonrelativistic calculations  
based on the distorted wave impulse approximation (DWIA) and with consistent
theoretical ingredients for bound and scattering states (i.e. overlap functions,
spectroscopic factors and optical model parameters able to give a good
description of ($e,e'p$) data) are unable to describe ($\gamma,p$)
data~\cite{Benenti,Ireland,Gaid}. A reasonable agreement and a consistent
description is obtained when the contribution of MEC is added to DKO in the 
($\gamma,p$) reaction~\cite{Gaid}. MEC produce a significant enhancement of the 
($\gamma,p$) cross sections calculated with DKO and affect both the shape and 
the magnitude of the angular distributions. For the ($\gamma,n$) reaction, 
where the DKO mechanism gives only a small fraction of the measured cross 
section, MEC and more complicated processes give the dominant 
contribution~\cite{Oxford,Benenti,Andersson}.   

However, the relative importance of the different mechanisms on 
($\gamma,p$) and ($\gamma,n$) reactions is still not completely understood and 
justifies the interest on other effects, like relativistic corrections, nuclear
current ambiguities and off-shell behavior of the bound nucleons.

The relativistic approach was first applied to ($\gamma,p$) reactions in 
Ref.~\cite{mcder}, where also MEC were considered, and in 
Refs.~\cite{lotz,Joha} within the framework of DKO. 
In these models the wave functions of the bound and continuum nucleons are 
solutions of a Dirac equation containing appropriate  scalar and vector 
potentials fitted to the ground state properties of the nucleus and to 
proton-nucleus elastic scattering data. The DKO mechanism was able to reproduce 
the $^{16}$O$(\gamma,p)$ cross section for an incident photon energy of 60 
MeV~\cite{Joha}. The same approach was then extended to several target nuclei 
and to a much wider energy range falling into the $\Delta$-excitation 
region~\cite{Joha1}. The comparison between these calculations and data suggests 
that DKO is the leading contribution for missing momentum values up to about 
500 MeV/$c$, while for larger values of the missing momentum an important effect
is expected from MEC and $\Delta$-excitation. 

Other studies within the same theoretical approach discussed the differences 
between relativistic and nonrelativistic calculations for $(\gamma,p)$ and 
($e,e'p)$ reactions~\cite{hedayati94,HPa}. They found noticeable medium 
modifications in the interaction hamiltonian due to relativistic potentials, 
which suggest that the role of MEC could be strongly modified with respect to
a nonrelativistic approach. In any case these relativistic models did not 
consider the ($\gamma,n$) reaction.   

Different models based on a fully relativistic DWIA (RDWIA) framework have
been developed in recent years and successfully applied to the analysis of
($e,e'p$) data~\cite{RDWIA,meucci}. In a recent paper~\cite{meucci} we have 
compared relativistic and nonrelativistic calculations for the ($e,e'p$) 
knockout 
reaction in order to study relativistic effects for cross sections and structure 
functions and to establish a limit in energy of the validity of a 
nonrelativistic approach. In this paper we make a similar comparison for 
($\gamma,N$) reactions. Relativistic effects are different in different 
situations and kinematics. In ($\gamma,N$) at intermediate photon energies the 
mismatch between the momentum transfer and the momentum of the outgoing nucleon 
is quite large and larger values of the missing momentum are explored than in 
usual ($e,e'p$) experiments. Thus, different effects can be expected for the two 
reactions. Our aim is to clarify the relationship between the RDWIA and DWIA 
approaches for ($\gamma,p$) and ($\gamma,n$) reactions also in comparison with 
data, and to check 
the relevance of the DKO mechanism in relativistic and nonrelativistic 
calculations. 

The RDWIA treatment is the same as in Ref.~\cite{meucci}. The relativistic 
bound state wave functions have been generated as solutions of a Dirac equation 
containing scalar and vector potentials obtained in the framework of the 
relativistic mean field theory.
The effective Pauli reduction has been adopted for the outgoing nucleon wave 
function. This scheme appears simpler and is in principle equivalent to the 
solution of
the Dirac equation. The resulting Schr\"odinger-like equation is
solved for each partial wave starting from relativistic optical potentials.
In the nonrelativistic calculations, the bound nucleon wave function has been 
taken as the normalized upper component of the relativistic four-component 
spinor and the scattering state is the solution of the same Schr\"odinger 
equivalent equation of the relativistic calculation.
In order to allow a consistent analysis of ($e,e'p$) and ($\gamma,p$) reactions
in comparison with data, RDWIA and DWIA calculations have been performed with 
the same bound state wave functions and optical potentials used for ($e,e'p$) 
in Ref.~\cite{meucci}. The same spectroscopic factors obtained in
Ref.~\cite{meucci} by fitting our RDWIA ($e,e'p$) results to data have been 
applied to the calculated ($\gamma,N$) cross sections. 

Results for $^{12}$C and $^{16}$O target nuclei at different photon energies
have been considered for the comparison. The relativistic current is written 
following the most commonly used current conserving $(cc)$  prescriptions 
for the ($e,e'p$) reaction introduced in Ref.~\cite{deF}. The ambiguities 
connected with different choices of the electromagnetic current cannot be 
dismissed. In the $(e,e'p)$ reaction the predictions  of different 
prescriptions are generally in close agreement~\cite{pollock}. Large differences
can however be found at high missing momenta~\cite{off1,off2}. 
These differences are expected to increase in ($\gamma,N$) reactions, where the
kinematics is deeply off-shell and higher values of the missing momentum 
are probed.

The formalism is outlined in Sec.~\ref{formalism}. Relativistic and
nonrelativistic calculations of the $^{12}$C$(\gamma,p)$ and
$^{16}$O$(\gamma,p)$ cross sections are compared in Sec.~\ref{rel}, where 
various relativistic effects and current ambiguities are investigated. In 
Sec.~\ref{neu} we discuss the role of the DKO mechanism in the description of
the $(\gamma,n)$ reaction. Some conclusions are drawn in Sec.~\ref{con}.

\section{Formalism}
\label{formalism}

The $(\gamma,N)$ differential cross section can be written as
\begin{eqnarray}
\sigma_{\gamma} = \frac{2\pi^2\alpha }{E_{\gamma}} \mid \p'\mid E'
 f_{\mathrm{rec}}f_{11} \ , \label{eq.unpgcs}
\end{eqnarray} 
where $E_{\gamma}$ is the incident photon energy, $E'$ and $\mid \p'\mid$ are 
the energy and the momentum of the emitted nucleon, and $f_{\mathrm{rec}}$ is
the recoil factor, which is given by 
\begin{eqnarray}
f_{\mathrm{rec}}^{-1} = 1 - \frac{E'}{E_{\mathrm{rec}}}
 \frac{\p'\cdot\p_{\mathrm{rec}}}{\mid\p'\mid^2} \ ,
\end{eqnarray}
where $E_{\mathrm{rec}}$ and $\p_{\mathrm{rec}}$ are the energy and the momentum
of the residual recoiling nucleus. 
In the cross section of Eq.~(\ref{eq.unpgcs}) only the transverse response,
$f_{11}$, appears.

If the photon beam is linearly polarized the cross section becomes
\begin{eqnarray}
\sigma_{\gamma,A} = \sigma_{\gamma} \left[ 1 + A\cos\left(2\phi\right)\right] \
, \label{eq.pgcs}
\end{eqnarray}
where $\phi$ is the angle between the photon polarization and the reaction
plane, and $A$ is the photon asymmetry, which can be expressed as the ratio
between the interference transverse-transverse and the pure transverse 
responses
\begin{eqnarray}
A = -\frac{f_{1-1}}{f_{11}} \ . \label{eq.A}
\end{eqnarray}
The structure functions $f_{\lambda \lambda'}$ are defined as bilinear
combinations of the nuclear current components, i.e.
\begin{eqnarray}
f_{11} &=& \langle J^x \left(J^x\right)^{\dagger} \rangle +
          \langle J^y \left(J^y\right)^{\dagger} \rangle \ , \nonumber \\
f_{1-1} &=&  \langle J^y \left(J^y\right)^{\dagger} \rangle -
          \langle J^x \left(J^x\right)^{\dagger} \rangle \ ,
\end{eqnarray}
where $\langle \dots \rangle$ means that average over the initial and sum over
the final states is performed fulfilling energy conservation. 
In our frame of reference the $z$ axis is along $\q$, and the $y$ axis is
parallel to $\q\times\p'$.

In  RDWIA the matrix elements of the nuclear current operator, i.e.  
\begin{eqnarray}
J^{\mu} = \int \diff \r \ove \Psi_f(\r) \hat j^{\mu} \exp{ \{ i\q\cdot \r\} }
 \Psi_i(\r) \ , \label{eq.rj}
 \end{eqnarray}
are calculated using relativistic wave functions for initial and final states.

The choice of the electromagnetic operator is, to some extent, arbitrary. Here
we discuss the three $cc$  expressions
\cite{deF,Kelly2,Kelly3}

\begin{eqnarray}
\hat j_{cc1}^{\mu} &=& G_M(Q^2) \gamma ^{\mu} - 
             \frac {\kappa}{2M} F_2(Q^2)\overline P^{\mu} \ , \nonumber \\
\hat j_{cc2}^{\mu} &=& F_1(Q^2) \gamma ^{\mu} + 
             i\frac {\kappa}{2M} F_2(Q^2)\sigma^{\mu\nu}q_{\nu} \ ,
	     \label{eq.cc} \\
\hat j_{cc3}^{\mu} &=& F_1(Q^2) \frac{\overline P^{\mu}}{2M} + 
             \frac {i}{2M} G_M(Q^2)\sigma^{\mu\nu}q_{\nu} \ , \nonumber
\end{eqnarray}
where $q^{\mu} = (\q,\omega)$ is the four momentum transfer,
$Q^2=\mid\q\mid^2-\omega ^2$, $\overline P^{\mu} = (E+E',\p+\p')$, 
$\kappa$ is the anomalous part of the magnetic
moment, $F_1$ and $F_2$ are the Dirac and Pauli nucleon form factors, $G_M =
F_1+\kappa F_2$ is the Sachs nucleon magnetic form factor, and
$\sigma^{\mu\nu}=i/2\left[\gamma^{\mu},\gamma^{\nu}\right]$.  Since the photon
is real, $Q^2=0$. In this case $F_1$ reduces to the nucleon total charge 
(1 for the proton, and 0 for the neutron), and $F_2$ to 1. Current 
conservation is restored by replacing the bound nucleon energy by \cite{deF}
\begin{eqnarray}
E = \sqrt{\mid \p \mid^2 + M^2} = \sqrt{ \mid \p'-\q\mid^2 + M^2} \ .
\end{eqnarray}

The bound state wave function
\begin{eqnarray}
\Psi_i = \left(\begin{array}{c} u_i \\ v_i \end{array}\right) \ , \label{eq.bwf}
\end{eqnarray}
is given by the Dirac-Hartree solution of a relativistic Lagrangian
containing scalar and vector potentials. 

The ejectile wave function $\Psi_f$ is written in terms of its positive energy
component $\Psi_{f+}$ following the direct Pauli reduction method
\begin{eqnarray}
\Psi_f = \left(\begin{array}{c} \Psi_{f+} \\ \frac {\ss\cdot\p'}{M+E'+S-V}
        \Psi_{f+} \end{array}\right) \ ,
\end{eqnarray}
where $S=S(r)$ and $V=V(r)$ are the scalar and vector potentials for the nucleon
with energy $E'$. The upper component $\Psi_{f+}$ can be related to a
Schr\"odinger equivalent wave function $\Phi_{f}$ by the Darwin factor $D(r)$,
i.e.
\begin{eqnarray}
\Psi_{f+} &=& \sqrt{D(r)}\Phi_{f} \ , \\
D(r) &=& \frac{M+E'+S-V}{M+E'} \ .
\end{eqnarray}
$\Phi_{f}$ is a two-component wave function which is solution of a 
Schr\"odinger
equation containing equivalent central and spin-orbit potentials obtained from
the scalar and vector potentials~\cite{HPa}. Hence, using the relativistic 
normalization, the emitted nucleon wave function is written as
\begin{eqnarray}
\ove \Psi _f = \Psi _f^{\dagger}\gamma ^0 &=& \sqrt {\frac {M+E'}{2E'}}\
\left[ \left(\begin{array}{c} 1 \\ \frac {\ss \cdot \p'}{C}\end{array} \right) 
\sqrt {D}\ \Phi _f \right] ^{\dagger }\ \gamma ^0  \nonumber \\
&=& \sqrt {\frac {M+E'}{2E'}}\ \Phi _f^{\dagger }
\left(\sqrt {D}\right) ^{\dagger } \left( 1\ ;\ \ss \cdot \p' 
\frac {1}{C^{\dagger }}\right) \gamma ^0  \ , \label{eq.psif}
\end{eqnarray}
where 
\begin{eqnarray}
C = C(r) = M + E' + S(r) - V(r) \ . \label{eq.cf}
\end{eqnarray}

If we substitute Eqs.~(\ref{eq.bwf}) and (\ref{eq.psif}) into
Eq.~(\ref{eq.rj}) and choose one of the current conserving prescriptions of 
Eq.~(\ref{eq.cc}), we obtain the relativistic expressions of the nuclear current
\begin{eqnarray}
\J_{cc1} &=& \sqrt {\frac {E'+M}{2E'}}\ \left\lmoustache
 \diff \r \ \Phi ^{\dagger }_f\ 
\left(\sqrt {D}\right) ^{\dagger } \left\{G_M \left[ \ss v_i
-i(\ss \cdot \dd)
\frac {1}{C^{\dagger }}\ \ss \ u_i \right]\right.\right.  \nonumber \\
&+& \frac {\kappa }{2M}F_2 \left.\left[ \left(2i\dd+\q\right) u_i + i 
(\ss \cdot \dd) \frac {1}{C^{\dagger }} \left(2i\dd+\q\right) 
  v_i \right]\right\} 
 \exp \{i\q \cdot \r \}\ ,  \label{eq.corcc1} \\
\J_{cc2} &=& \sqrt {\frac {E'+M}{2E'}}\ \left\lmoustache
 \diff \r \ \Phi ^{\dagger }_f\ 
\left(\sqrt {D}\right) ^{\dagger } \left\{F_1 \left[ 
-i(\ss \cdot \dd)
\frac {1}{C^{\dagger }}\ \ss \ u_i + \ss \ v_i\right]\right.\right.
\nonumber  \\
&+& i\frac {\kappa }{2M}F_2 \left[ \ss \times \q  u_i + \omega 
(\ss \cdot \dd) \frac {1}{C^{\dagger }} \ss  u_i \right. \nonumber \\
&-& \left.\left. i\omega \ss  v_i + i(\ss \cdot \dd)
\frac {1}{C^{\dagger }} \ss \times \q \ v_i\right]
\right\} \exp \{i\q \cdot \r \}\ , \label{eq.corcc2}  \\
\J_{cc3} &=& \sqrt {\frac {E'+M}{2E'}}\ \left\lmoustache
 \diff \r \ \Phi ^{\dagger }_f\ 
\left(\sqrt {D}\right) ^{\dagger }\right. \nonumber \\
&& \left\{\frac {i}{2M}F_1 \left[ 
\left(-2i\dd-\q\right) u_i - i\left( \ss \cdot \dd \right)
\frac {1}{C^{\dagger }} \left(2i\dd+\q\right) v_i \right]\right. \nonumber \\
&+& \frac {i}{2M}G_M \left[ \ss \times \q  u_i + \omega 
(\ss \cdot \dd) \frac {1}{C^{\dagger }} \ss  u_i\right. \nonumber \\  
&-& \left.\left.i\omega \ss  v_i + i(\ss \cdot \dd)
\frac {1}{C^{\dagger }} \ss \times \q  v_i\right]
\right\} \exp \{i\q \cdot \r \}\ ,  \label{eq.corcc3} 
\end{eqnarray}
where the $\overline\P $ operator has been replaced by the gradient $-2i\dd-\q$,
which operates not only on the components of the Dirac spinor but also on $\exp
\{i\q \cdot \r \}$. It is interesting to notice that in Eqs.~(\ref{eq.corcc1})
and (\ref{eq.corcc3}) appear terms which
are proportional to the second derivative of the lower component of the Dirac
spinor.

\section{The $(\gamma,\lowercase {p})$ reaction}
\label{rel}

The $(\gamma,p)$ reaction is an interesting process for testing our RDWIA
program and investigating the differences with respect to the DWIA approach. 
At intermediate photon energies there is a large difference 
between the incoming photon and outgoing nucleon momenta and missing momentum 
values higher than in usual ($e,e'p$) experiments are explored. 
Thus, different relativistic
effects can be expected in the two reactions. Moreover, it can be interesting to
check the relevance of the DKO mechanism in comparison with data for
corresponding RDWIA and DWIA calculations with consistent theoretical
ingredients for bound and scattering states. Previous RDWIA
analyses~\cite{Joha1} suggest 
that DKO is the leading contribution to the ($\gamma,p$) cross section for 
low values of $E_\gamma$ and not too large values of the missing momentum. In
contrast, in nonrelativistic calculations the DKO mechanism generally 
underestimates the experimental cross sections and an important contribution is
given by MEC even at low photon energies. In these investigations, however, 
RDWIA and DWIA calculations make generally use of different bound state wave 
functions and optical potentials, and ($\gamma,p$) results are very sensitive to 
the theoretical ingredients adopted in the calculations. 

A large amount of experiments were carried out in the past on several target
nuclei and over a wide range of photon energies. Here, we have performed 
calculations for $^{12}$C and $^{16}$O. The bound state wave functions and 
optical potentials are the same as in the analysis of Ref.~\cite{meucci}, 
where the RDWIA results are in satisfactory agreement with ($e,e'p$) data. 
In order to allow a consistent comparison with data, the same spectroscopic 
factors obtained by fitting our RDWIA $(e,e'p)$ calculations~\cite{meucci}
to data have been here applied to the ($\gamma,p$) results, that is 0.56 for 
$^{12}$C and 0.70 for $^{16}$O. 

The relativistic bound state wave function has been generated using the program
ADFX of Ref.~\cite{adfx}, where relativistic Hartree-Bogoliubov equations are
solved. The model starts from a Lagrangian density containing sigma-, omega-,
rho-meson, and photon fields, whose potentials are obtained by solving
self-consistently Klein-Gordon equations.

The corresponding wave function for the nonrelativistic calculation has been
taken as the upper component of the relativistic four-component spinor, which is
normalized to 1 in coordinate and spin space. Presumably, this is not the best
choice for the nonrelativistic DWIA calculations,  but the same ingredients are 
to be used in order to perform a clear comparison between the two approaches.

The outgoing nucleon wave function is calculated by means of the complex
phenomenological optical potential of Ref.~\cite{chc}, obtained from fits to
proton elastic scattering data in an energy range up to 1040 MeV. The
Schr\"odinger equivalent potentials calculated in the same way were used in the
nonrelativistic program.

Since no rigorous prescription exists for handling off-shell nucleons,
it is worthwhile to study the sensitivity of one nucleon photoemission to
different choices of the nuclear current.

The nonrelativistic current is written as an expansion up to order $1/M^2$ from 
a Foldy-Wouthuysen transformation~\cite{FW,giusti80} applied to the interaction
Hamiltonian where the nuclear current is in  the $cc2$ form of Eq.~(\ref{eq.cc}).
Thus, the $cc2$ prescription for the relativistic nuclear current is more 
appropriate in the comparison between the relativistic and nonrelativistic 
models.

\subsection{Relativistic and nonrelativistic calculations}

In this section the results of the comparison between our RDWIA and DWIA
calculations are discussed. One has to remember that our
nonrelativistic code contains some relativistic corrections in the kinematics
and in the nuclear current through the expansion in $1/M$. This means that the
nonrelativistic results cannot be obtained from the relativistic program simply 
by neglecting the lower components of the Dirac spinor and applying the proper
normalization.

The comparison between the RDWIA and DWIA results is shown in Fig.~\ref{f.cso} 
for the cross section of the $^{16}$O$(\gamma,p)^{15}$N$_{\mathrm {g.s.}}$ 
reaction. The photon energy range is taken between 60 MeV and 257 MeV, but the 
nonrelativistic calculations are not extended above 200 MeV \cite{meucci}. In
the considered energy range missing momentum values between about 200 and 1000
MeV/$c$ are explored.  

We see that the differences between the nonrelativistic calculations and the
relativistic ones with the $cc2$ prescription are sensible at all energies. 
The nonrelativistic results are always smaller than the
data~\cite{miller,findlay,leitch,adams}. This effect was 
already known from previous nonrelativistic analyses and suggested that MEC must
give an important contribution to the cross section. On the contrary, the 
relativistic results are generally closer to the data and well reproduce the 
magnitude and shape, at least at low energies. This result is in agreement 
with similar RDWIA approaches with the $cc2$ current~\cite{lotz,Joha,Joha1}.   
For higher energies, the relativistic results fall below the data and the 
discrepancies increase with the proton angle. This seems to indicate that the 
DKO mechanism gives the most important contribution to the cross section at 
lower missing momenta, while more complicated processes such as MEC and 
$\Delta$-excitations become more and more important at larger missing momenta.

In Fig.~\ref{f.ao} the photon asymmetries are shown in the same kinematics as 
in Fig.~\ref{f.cso}. The differences between DWIA and RDWIA results with $cc2$ 
are small at 60 MeV, but rapidly increase with the photon energy. 

In Fig.~\ref{f.csc} the cross section for the 
$^{12}$C$(\gamma,p)^{11}$B$_{\mathrm{g.s.}}$ reaction is presented. The 
nonrelativistic results are also in this case smaller than the relativistic 
ones, but the most apparent feature is that both results lie above the
data~\cite{spring,shotter}.  
The fact that RDWIA calculations with the $cc2$ current
overestimate the data by a factor of 2 was already pointed out in
Ref.~\cite{Joha1}. A better description of data might be obtained with a more
careful determination of the $^{12}$C ground state which should 
include its intrinsic deformation.

\subsection{Current ambiguities}

In this section the sensitivity of $(\gamma,p)$ calculations to different
choices of the nuclear current is discussed. In the case of one proton knockout 
the expressions for the electromagnetic nuclear current of
Eq.~(\ref{eq.cc}) reduce to
\begin{eqnarray}
\hat j_{cc1}^{\mu} &=&  \gamma ^{\mu} + 
             \kappa_p \left(\gamma ^{\mu} - \frac{\overline P^{\mu}}{2M}\right)
	      \ , \nonumber \\
\hat j_{cc2}^{\mu} &=& \gamma ^{\mu} + 
             i\frac {\kappa_p}{2M}\sigma^{\mu\nu}q_{\nu} \ ,
	     \label{eq.ccp} \\
\hat j_{cc3}^{\mu} &=& \frac{\overline P^{\mu}}{2M} + 
             \frac {i}{2M} \left(1+\kappa_p\right)\sigma^{\mu\nu}q_{\nu} 
	     \ , \nonumber
\end{eqnarray}
where $\kappa_p = 1.793$ is the anomalous part of the proton magnetic
moment. These expressions are obviously equivalent for a free nucleon, but give
different results for an off-shell nucleon.

It is interesting to notice that the nonrelativistic reductions of the three
$cc$ forms give identical results up to order $1/M$ following the
direct Pauli reduction scheme in the limit of no Dirac $S$ and $V$ potentials
and $M+E=2M$. The equivalence of Pauli reduction and Foldy-Wouthuysen
transformation up to order $1/M$ was already pointed out in
Refs.~\cite{fearing,HPa}. 

\begin{figure}[h]
\begin{center}
\includegraphics[height=20cm]{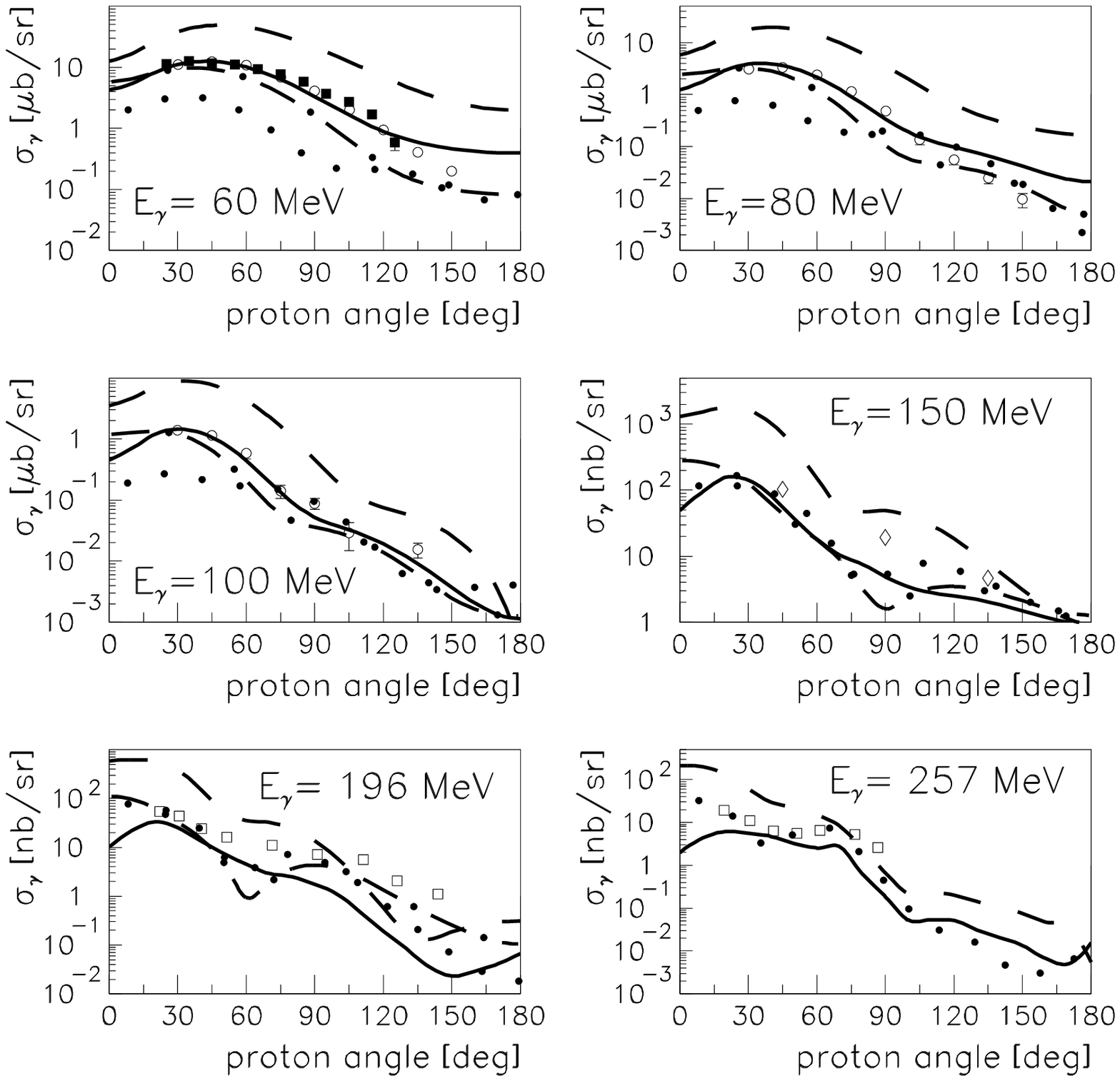}
\vskip -1.5cm
\end{center}
\caption{The cross section for the $^{16}$O$(\gamma,p)^{15}$N$_{\mathrm
{g.s.}}$ reaction as a function of the proton scattering angle 
for photon energies ranging from 60 to 257 MeV.
The data at 60 MeV are from
Ref.~\cite{miller} (black squares) and from 
Ref.~\cite{findlay} (open circles). The data at 80 and 100 MeV are from
Ref.~\cite{findlay}. The data at 150 MeV are from
Ref.~\cite{leitch}, and those at 196 and 257 MeV are
from Ref.~\cite{adams}. Results shown correspond to 
RDWIA calculations with the $cc2$ (solid line), $cc1$ (dashed line), and $cc3$
(dotted line) current. The dot-dashed line is the nonrelativistic result.  
\label{f.cso}}
%\end{center}
\end{figure}

\begin{figure*}[ht]
\begin{center}
\includegraphics[height=20cm]{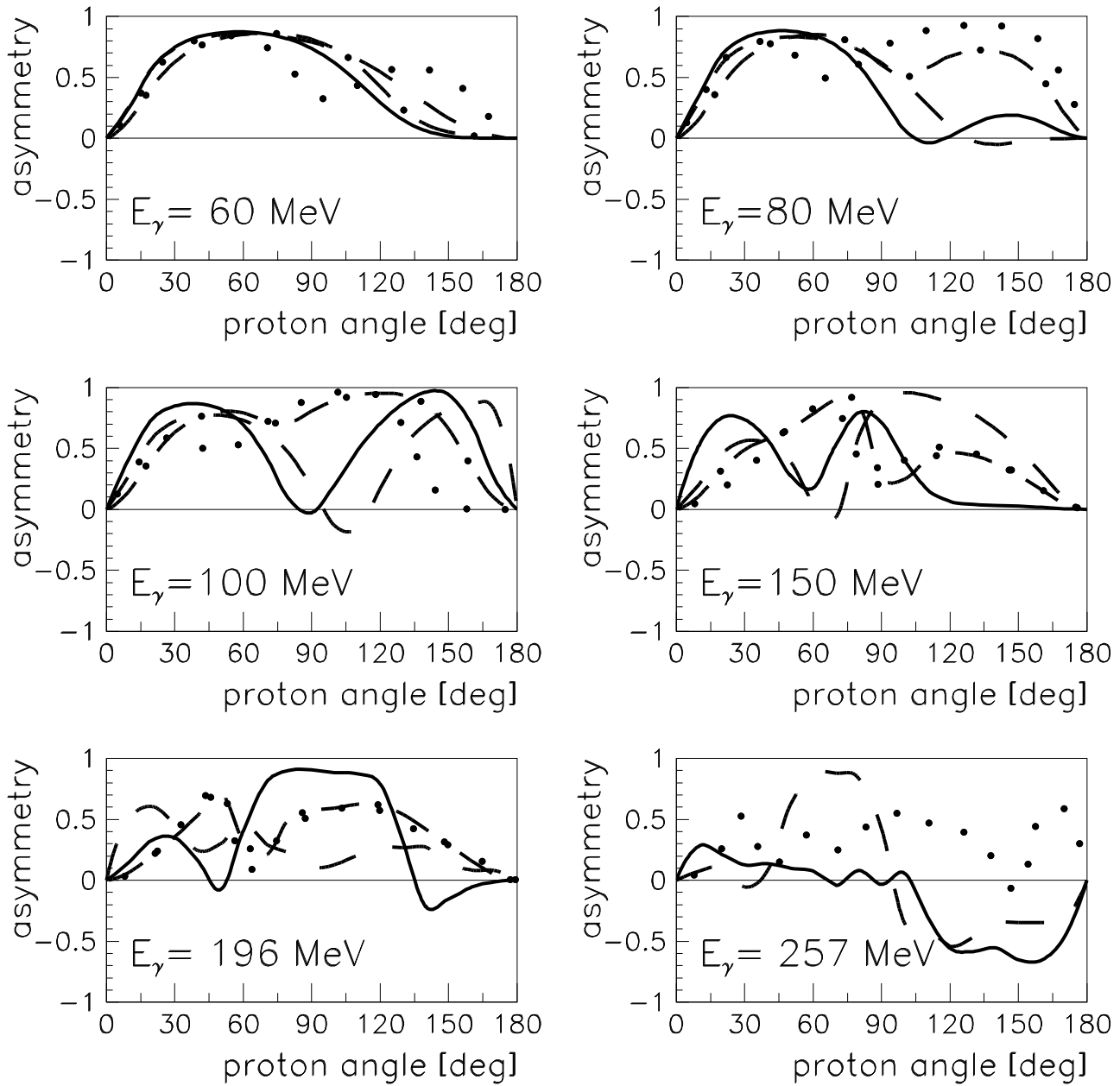}
\vskip -1.5cm
\caption{The same as in Fig.~\ref{f.cso}, but for the photon asymmetry. 
\label{f.ao}}
\end{center}
\end{figure*}

\begin{figure}[ht]
\begin {center}
\includegraphics[height=20cm]{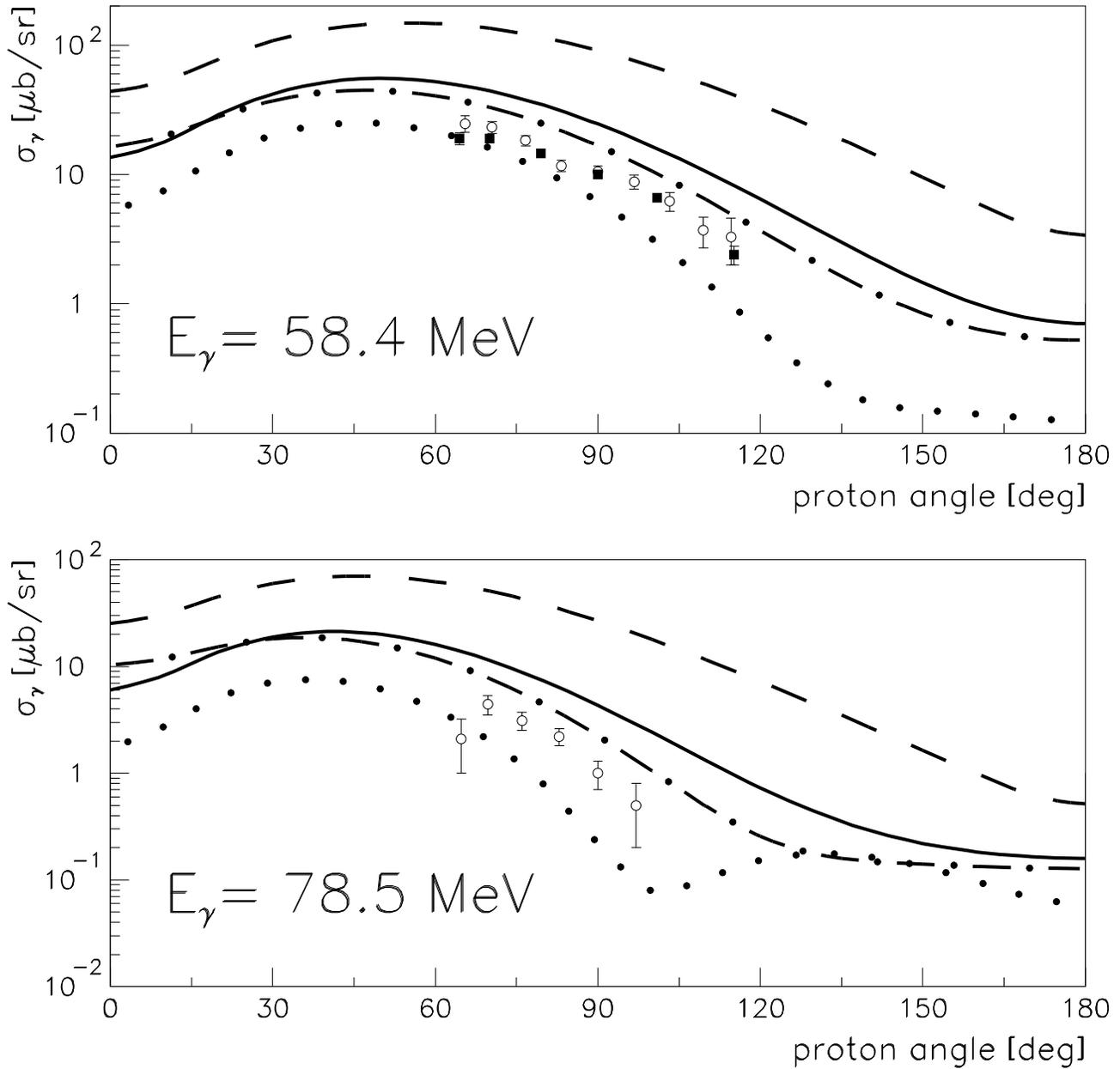}
\vskip -1.5cm
\caption {The cross section for the $^{12}$C$(\gamma,p)^{11}$B$_{\mathrm
{g.s.}}$ reaction as a function of the proton scattering angle 
at $E_{\gamma}$ = 58.4 and 78.5 MeV. The data are from
Ref.~\cite{shotter} (black squares) and from Ref.~\cite{spring} (open circles). 
Line convention as in Fig.~\ref{f.cso}. 
\label{f.csc}}
\end {center}
\end{figure}

The results obtained with different current operators are
displayed and compared for $^{16}$O in Fig.~\ref{f.cso}. 
The differences are large. We have already noticed that the $cc2$ 
results are in satisfactory agreement with the experimental data at lower 
energies, but they tend to fall down with increasing proton angle and photon
energy. RDWIA results are strongly enhanced if we use $cc1$
current. This is probably due to a too small interference term which does not 
correctly estimate the convective current contained in both  
$\gamma^{\mu}$ and $\overline P^{\mu}/(2M)$ terms when the nucleon is off-shell.
Also in Ref.~\cite{off1}, in an ($e,e'p$) analysis within the framework of the
relativistic plane wave impulse approximation, large differences are found
between results obtained with the $cc2$ and $cc1$ prescriptions for
high values of the missing momentum and significantly higher cross sections are
obtained with $cc1$. 
The results with the $cc3$ current in Fig.~\ref{f.cso} are more similar to the 
$cc2$ ones. At low energy $cc3$ lies below $cc2$, but the differences
rapidly decrease with the energy.

In Fig.~\ref{f.ao} a comparison of photon asymmetry calculations in the same
kinematics as in Fig.~\ref{f.cso} is shown. The differences are sensible 
already at 60 MeV and tend to increase with the energy.

Large ambiguities are found also in the case of $^{12}$C($\gamma,p$) reaction 
(Fig.~\ref{f.csc}). Results obtained with the $cc1$ current are enhanced 
above the data by an order of magnitude. On the contrary, $cc3$ results are 
smaller than the data.

\subsection{Spinor distortion and Darwin factor}

The optical potential enters into the Darwin factor $D$,
which multiplies the Schr\"odinger equivalent eigenfunction, and into the spinor
distortion $C$, which is applied only to the lower component of Dirac spinor.
The distortion of the scattering wave function is calculated through a partial 
wave expansion and it is always included in the calculations. The Darwin factor
gives a reduction of the cross section. On the contrary, the spinor distortion 
produces an enhancement. 

\begin{figure}[ht]
\begin {center}
\includegraphics[height=20cm]{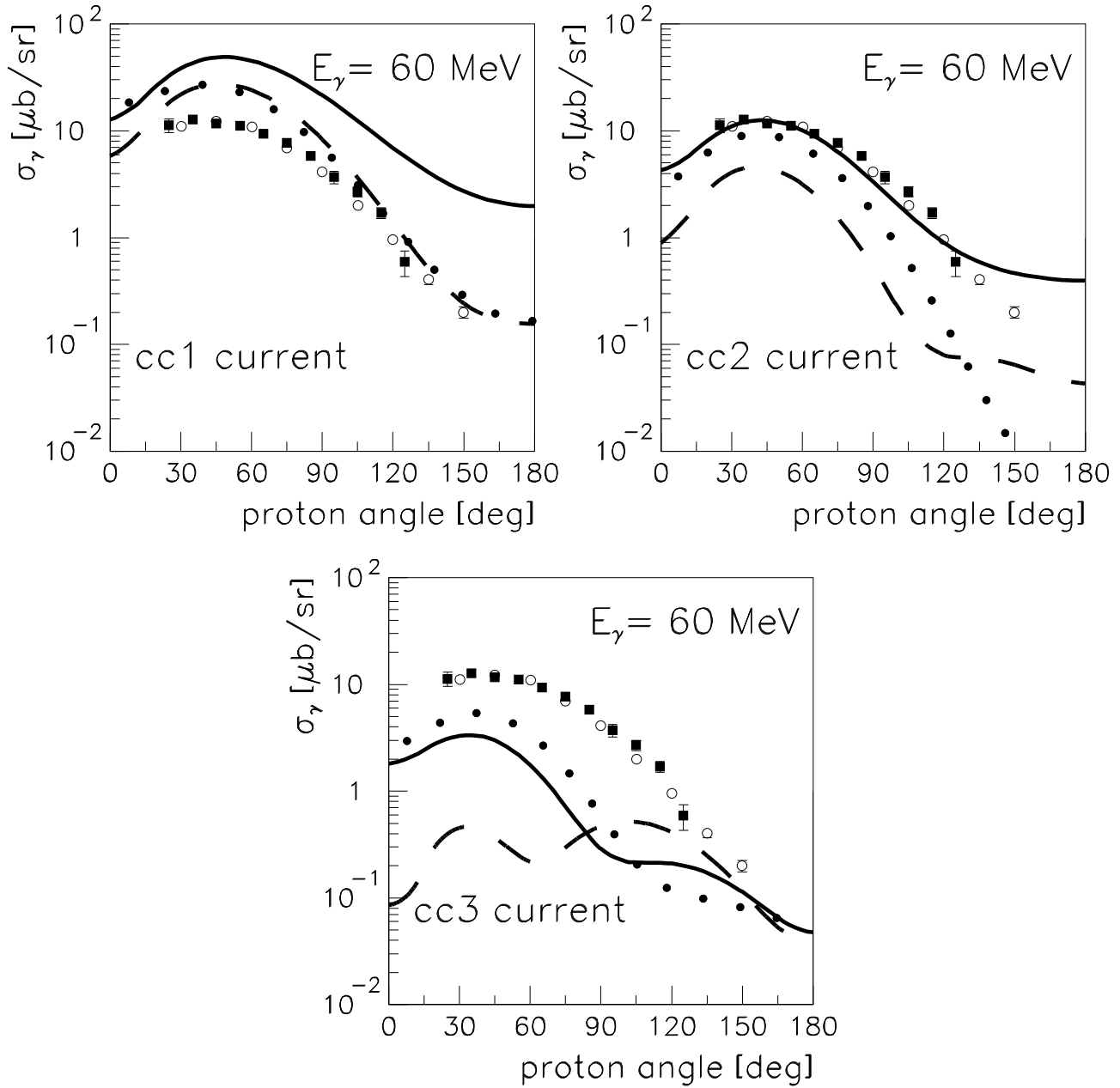}
\vskip -1.5cm
\caption {The cross section for the $^{16}$O$(\gamma,p)^{15}$N$_{\mathrm
{g.s.}}$ reaction as a function of the proton scattering angle 
at $E_{\gamma}$ = 60 MeV. The data are from
Ref.~\cite{miller} (black squares) and from Ref.~\cite{findlay} (open circles). 
The solid lines give the RDWIA results, the
dotted lines the calculations without the Darwin factor and spinor distortion, 
and the dashed lines the EMA. 
\label{f.sve}}
\end {center}
\end{figure}

\begin{figure}[ht]
\begin {center}
\includegraphics[height=20cm]{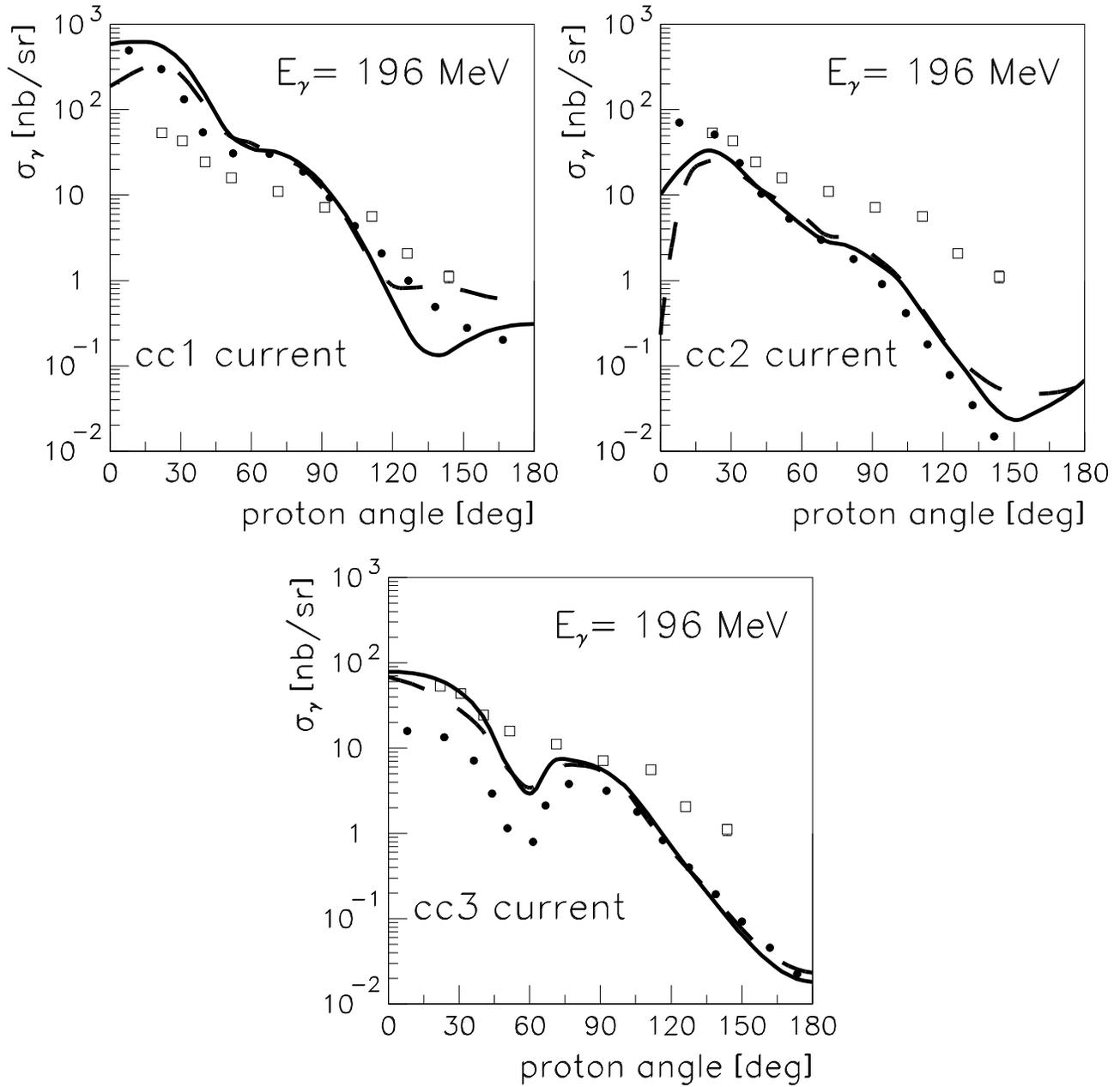}
\vskip -1.5cm
\caption {The same as in Fig.~\ref{f.sve}, but at $E_{\gamma}$ = 196 MeV. The
data are from Ref.~\cite{adams}.
\label{f.sve1}}
\end {center}
\end{figure}

The combined effects of the two corrections are displayed and
compared in Figs.~\ref{f.sve} and \ref{f.sve1} for the cross section of the 
reaction $^{16}$O$(\gamma,p)^{15}$N$_{\mathrm {g.s.}}$ 
at $E_{\gamma}$ = 60 and 196 MeV. Results without the Darwin factor and 
spinor distortion at 60 MeV using either $cc1$ or $cc2$  are
reduced with respect to the full calculations, while results with $cc3$ are 
enhanced for low scattering angles. These effects decrease at 196 
MeV, where calculations without potentials are closer to full calculations.

\subsection{Effective momentum approximation}

The EMA prescription, which consists in evaluating the momentum operator in the
nuclear current using the asymptotic value of the ejected nucleon momentum, 
strongly simplifies the calculations. This approximation
was successfully used in some $(e,e'p)$ calculations, and, in particular, in the
model of Refs.~\cite{Kelly2,Kelly3} for bound and scattering states. Since in
our approach the bound state wave function is solution of a Dirac equation, we
investigate the EMA effects only for the scattering states. We have to notice
that in the nuclear current the EMA prescription affects only the $\overline
P^{\mu}$ term in $cc1$ and $cc3$ formulae, while $cc2$ is unchanged. However, a 
momentum dependence comes from the Pauli reduction of the scattering wave
function.

The effects of EMA are displayed and compared with the full RDWIA results in 
Figs.~\ref{f.sve} and \ref{f.sve1} at $E_{\gamma}$ = 60 and 196 MeV. At 60 MeV 
the differences are large, but they decrease with the energy and become much
smaller at 196 MeV. This behavior is practically independent of the nuclear
current. This can be understood if we consider that distortion effects decrease
with the energy, so that at high energy DWIA results are more similar to PWIA
ones, where EMA is exact.

\section{The $(\gamma,\lowercase {n})$ reaction}
\label{neu}

In this section relativistic effects are discussed for the ($\gamma,n$)
reaction. The experimental angular distributions are similar in magnitude and 
shape to those obtained for the $(\gamma,p)$ reaction. The ratio between the 
$(\gamma,p)$ and $(\gamma,n)$ cross sections is comparable to unity and suggests 
a two-body mechanism. In fact, nonrelativistic calculations based on the DKO
mechanism give but a small fraction of the measured cross sections.  

In order to test the relevance of the DKO contribution, we have 
performed RDWIA and DWIA calculations for the 
$^{16}$O$(\gamma,n)^{15}$O$_{\mathrm{g.s.}}$ reaction. For neutron 
knockout the electromagnetic nuclear current of Eq.~(\ref{eq.cc}) reduces to the 
anomalous spin current only, i.e.
\begin{eqnarray}
\hat j_{cc1}^{\mu} &=&   
             \kappa_n \left(\gamma ^{\mu} - \frac{\overline P^{\mu}}{2M}\right)
	     \ \ \ , \nonumber \\
\hat j_{cc2}^{\mu} &=& \hat j_{cc3}^{\mu}\ = \
             i\frac {\kappa_n}{2M}\sigma^{\mu\nu}q_{\nu} \ ,
	     \label{eq.ccn}
\end{eqnarray}
where $\kappa_n = -1.913$ is the anomalous part of the neutron magnetic
moment. Notice that $\hat j_{cc2}^{\mu} = \hat j_{cc3}^{\mu}$, while for $cc1$
the spin current is written by means of a difference between the Dirac current
$\gamma^{\mu}$ and the convective current $\overline P^{\mu}/(2M)$. 

\begin{figure}[ht]
\begin {center}
\includegraphics[height=20cm]{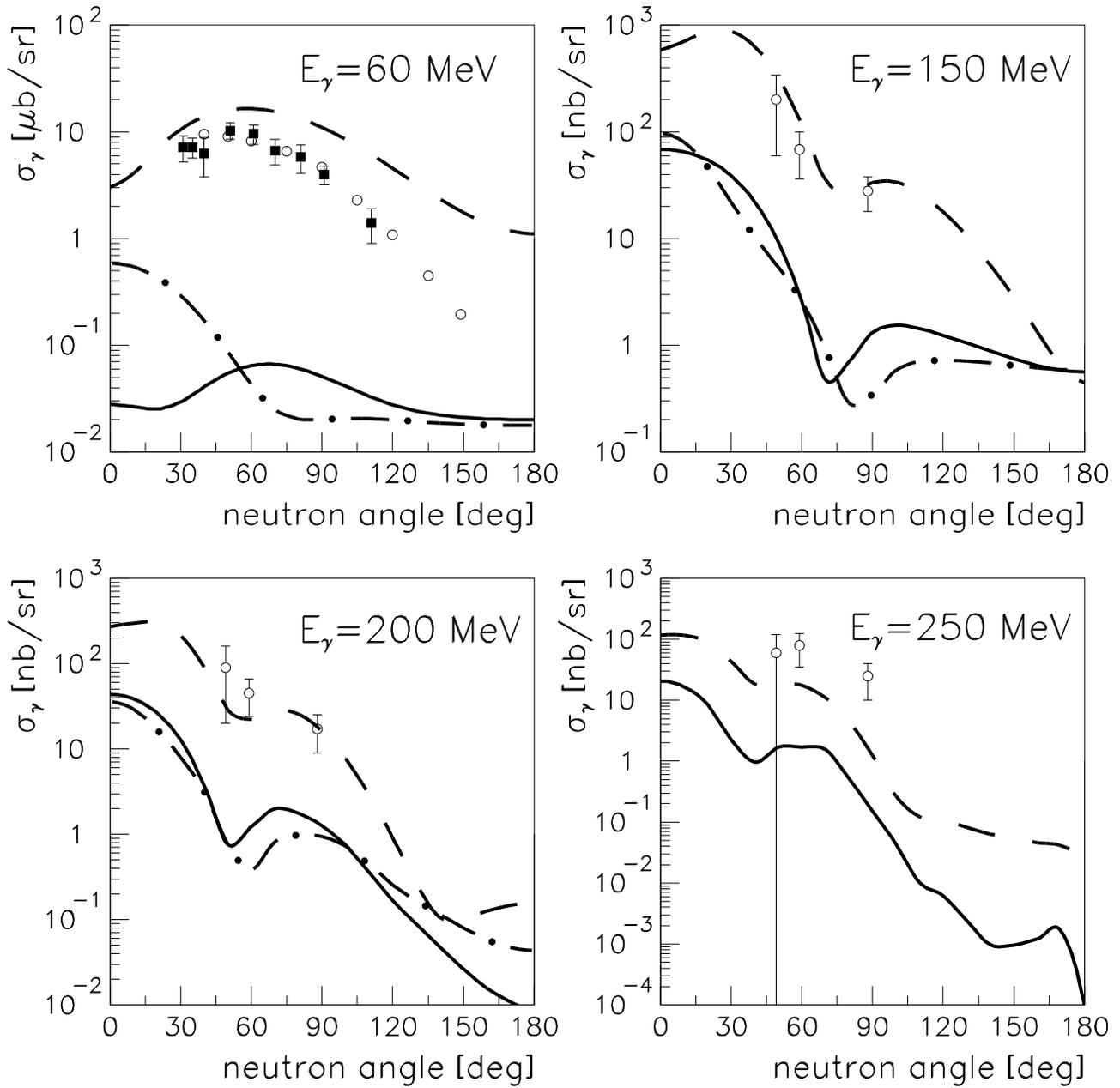}
\vskip -1.5cm
\caption {The cross section for the $^{16}$O$(\gamma,n)^{15}$O$_{\mathrm
{g.s.}}$ reaction as a function of the neutron scattering angle 
 for photon energies ranging from 60 to 250 MeV. The data at 60 MeV
are from Ref.~\cite{Andersson} (black squares) and from Ref.~\cite{goringer} 
(open circles), and the data at 150, 200, and 250 MeV are from
Ref.~\cite{beise}. Line convention as in Fig.~\ref{f.cso}. 
\label{f.cso_n}}
\end {center}
\end{figure}

In Fig.~\ref{f.cso_n} relativistic and nonrelativistic 
results for the $^{16}$O$(\gamma,n)^{15}$O$_{\mathrm{g.s.}}$ reaction are shown
in comparison with data~\cite{Andersson,goringer,beise}. The same spectroscopic 
factor as in the corresponding $(\gamma,p)$ reaction has been applied to the
calculated results.

We see that neither nonrelativistic nor relativistic $cc2$ $(cc3)$ calculations
reproduce the magnitude of experimental data. This result is not surprising. It
confirms what was already found in previous DWIA calculations and indicates that
more complicated two-body effects are needed to reproduce the data.
Relativistic results are strongly enhanced if we use the $cc1$ current. This 
effect is particularly surprising at $E_{\gamma} =$ 150 and 200
MeV, where the $cc1$ curve fits the data. This result can be attributed to the 
$\gamma^{\mu} - \overline P^{\mu}/(2M)$ operator, which does not correctly 
describe the spin current when the kinematics is deeply off-shell, and, 
therefore, is to be considered unreliable.

The differences between the DWIA and RDWIA results with $cc2$ are 
large. They are reduced if we perform nonrelativistic calculations with a
nuclear current expanded up to order $1/M^3$~\cite{giusti80}, but the 
contribution of the third order is very large for this reaction, and comparable 
to the second order one.

\section{Summary and conclusions}
\label{con}

In this paper we have presented relativistic and nonrelativistic DWIA 
calculations for $(\gamma,N)$ reactions on $^{12}$C and $^{16}$O, in a
photon-energy range between 60 and 257 MeV, in order to check the 
relevance of the DKO mechanism in RDWIA and DWIA models and investigate 
relativistic effects. 

The transition matrix element of the nuclear current operator is calculated in 
RDWIA using the bound state wave functions obtained in the framework of the 
relativistic mean field theory, and the direct Pauli reduction method with 
scalar and vector potentials for the ejectile wave functions. In order to study 
the ambiguities in the electromagnetic vertex due to the off-shellness of the 
initial nucleon, we have performed calculations using three current conserving 
expressions. The nonrelativistic DWIA matrix elements are computed in a similar 
way to allow a direct comparison with the relativistic results. In order to
allow a consistent comparison of ($e,e'p$) and ($\gamma,p$) data, calculations
have been performed with the same bound state wave functions, spectroscopic
factors and optical potentials as in our recent ($e,e'p$) analysis of 
Ref.~\cite{meucci}. 

Nonrelativistic $(\gamma,p)$ results are always smaller than the data and 
suggest the idea that MEC are relevant even at low energies. On the contrary, 
RDWIA calculations seem to indicate that the DKO mechanism is the leading 
process, at least for low photon energies. These results are in substantial 
agreement with previous DWIA and RDWIA analyses.

We have discussed the sensitivity of the $(\gamma,p)$ reaction to the different 
choices of the nuclear current. Unlike the case of the $(e,e'p)$ reaction, 
large ambiguities are generally found. Results with the $cc2$ current
are in satisfactory agreement with the experimental data at lower 
energies, but they tend to fall down with increasing proton angle and photon
energy. On the contrary, the results with $cc1$ are strongly enhanced. This 
result seems due to a too small interference term which overestimates the 
convective current when the initial nucleon is off-shell. 
The results with $cc3$  are more similar to the $cc2$ ones. The differences 
decrease when the energy increases.

The effect of the scalar and vector potentials in the Pauli reduction
for the scattering state has been discussed. These potentials appear in the 
relativistic treatment and are absent in the nonrelativistic one.
The combined contribution of the Darwin factor, which reduces the cross section,
and of the spinor distortion, which enhances the effects of the lower 
components of the Dirac spinor, is important at low $E_{\gamma}$, and decreases
at higher energies.

The validity of EMA in the scattering state of relativistic calculations 
has been investigated. The differences with respect to the exact result are 
large at low photon energies, but rapidly decrease and become small at higher 
energies.

Relativistic calculations of the $(\gamma,n)$ cross sections give huge off-shell 
ambiguities. The $cc2$ and $cc3$ prescriptions coincide in the neutron case, but
the enhancement obtained with $cc1$ is dramatic and brings the RDWIA results
above the data at $E_{\gamma}$ = 60 MeV and in good agreement with data at 
$E_{\gamma}$ = 150 and 200 MeV. However, we cannot argue that the DKO mechanism 
with the $cc1$ prescription correctly describes $(\gamma,n)$ cross sections. 
This result is due to a dominant off-shell effect on the $cc1$ current 
operator, which does not correctly describe the modest contribution from the 
spin current.

Neither norelativistic DWIA nor RDWIA calculations with $cc2$ reproduce
($\gamma,n$) data. There are sensible differences between the results of the 
two approaches, but in both cases the experimental cross sections are largely
underestimated. This is an indication of the dominance of two-body mechanisms 
in the ($\gamma,n$) reaction. A careful and consistent evaluation of these 
mechanisms within relativistic and nonrelativistic frameworks for ($\gamma,n$) 
and ($\gamma,p$) reactions would be highly desirable and helpful to 
draw conclusions about the reaction mechanism and to solve the present 
ambiguities.

%\newpage
%%%%%%%%
%%%%%%%%%%%%%%
%%
%% FIGURES
%%
%%%%%%%%%%%%%%
%%%%%%%%%%%%%%
 
%
%%%%%%%%%%%%%%%%%%%%%%%%%%%%%%%%%%%%%%%%%%%%%%%%%%%%%%%%%%%%%%%%%%%%%%%%%%%%%%%
%
%
%%%%%%%%%%%%%%%%%%%%%%%%%%%%%%%%%%%%%%%%%%%%%%%%%%%%%%%%%%%%%%%%%%%%%%%%%%%%%%%
%
%
%
%%%%%%%%%%%%%%%%%%%%%%%%%%%%%%%%%%%%%%%%%%%%%%%%%%%%%%%%%%%%%%%%%%%%%%%%%%%%%%%
%
%
%%%%%%%%%%%%%%%%%%%%%%%%%%%%%%%%%%%%%%%%%%%%%%%%%%%%%%%%%%%%%%%%%%%%%%%%%%%%%%%
%
%
%
%%%%%%%%%%%%%%%%%%%%%%%%%%%%%%%%%%%%%%%%%%%%%%%%%%%%%%%%%%%%%%%%%%%%%%%%%%%%%%%
%
%
%
%%%%%%%%%%%%%%%%%%%%%%%%%%%%%%%%%%%%%%%%%%%%%%%%%%%%%%%%%%%%%%%%%%%%%%%%%%%%%%%
%
%

%%%%%%%%%%%%%%%%%%%%%%%%%%%%%%%%%%%%%%%%

\end{document}